\documentclass[12pt,a4paper]{article}
\newcommand\be{\begin{equation}}
\newcommand\ee{\end{equation}}
\newcommand\bea{\begin{eqnarray}}
\newcommand\eea{\end{eqnarray}}
\newcommand\ket[1]{|#1\rangle}
\newcommand\bra[1]{\langle #1|}

\newcommand{\fatalpha}{{\bf \alpha \kern -0.44em \alpha}}
\newcommand{\fatsigma}{{\bf \sigma \kern -0.54em \sigma}}
\newcommand{\tpchi}{{\bf \chi \kern -0.35em \chi}}
\newcommand{\llambda}{{\bf \lambda \kern -0.45em \lambda}}

\begin{document}
\title{Quantum Diagonalization of Hermitean Matrices}
\author{Stefan Weigert \\
Institut de Physique, Universit\'e de Neuch\^atel\\
Rue A.-L. Breguet 1, CH-2000 Neuch\^atel, Switzerland\\
\tt stefan.weigert@iph.unine.ch}
\date{January 2000}
\maketitle
\begin{abstract}
To measure an observable
of a quantum mechanical system
leaves it in one of its eigenstates
and the result of the measurement is one of its eigenvalues.
This process is shown to be a {\em computational resource}. It
allows one, in principle, to diagonalize hermitean $(N \times N)$
matrices by quantum mechanical measurements only. To do so, one
considers the given matrix as an observable of a single spin with
appropriate length $s$
which can be measured using a generalized Stern-Gerlach apparatus.
Then, each run provides one  eigenvalue of the observable.
As it is based on the `collapse of the wave function' associated with
a measurement, the procedure is neither a digital nor an
analog calculation---it defines thus a new {\em quantum mechanical}
method of computation.
\end{abstract}
\vspace{8mm}
%
%
%
Non-classical features of quantum mechanics such as Heisenberg's
uncertainty relation and entanglement have intrigued physicists
for several decades. From a classical point of view, quantum
mechanics imposes constraints on the ways to talk about nature. An
electron does not ``have'' position and momentum as does a
billiard ball. Similarly, if a photon is entangled with a second
one---possibly very far away---one cannot ascribe properties to it
as is done for an individual classical particle. The lesson to be
learned is that classical intuition about the macroscopic world
simply does not extrapolate into the microscopic world.

In recent years, an entirely different attitude towards quantum theory
has been put forward. The focus is no longer on attempts to come to
terms with its strange features but to capitalize on its both
counter-intuitive and well-established
properties. In this way, surprising methods have been uncovered to
solve specific problems by means which have no classical
equivalent: {\em quantum cryptography}, for example, allows one to
establish secure keys for secret transmission of information
\cite{bennet+84}; entanglement \cite{schroedinger35/2} is used as
a tool to set up powerful {\em quantum algorithms} which do factor
large integers much more efficiently than any classical algorithm
\cite{shor94}. Throughout, these new techniques rely on the {\em
measurement} of quantum mechanical observables as a reliable tool.
This is also true for {\em quantum error correction}
\cite{shor95,steane96} required to let any potential algorithm
run.

Here the purpose is to point out that the bare `projection'
\cite{schroedinger35/2} effected by a quantum mechanical
measurement does possess computational power itself. As will be
shown below, it can be used to solve explicitly at least one
specific computational task, namely to determine eigenstates and
eigenvalues of hermitean $(N \times N)$ matrices.
%
%
%

%
%
The diagonalization of hermitean matrices is a recurrent problem
in mathematics, physics, and related fields. Using the notation of
a quantum physicist the problem reads as follows. Given a
self-adjoint operator ${\widehat A}$ acting on a Hilbert space
${\cal H}$ of dimension $N$, one needs to determine its
eigenstates $\ket{A_n}, n=1, \ldots, N,$ and its $N$ real
eigenvalues $A_n$ satisfying
%
%
${\widehat A} \, \ket{A_n} = A_n \ket{A_n}  ,  n=1, \ldots, N.$
%
%
If normalized to one, the eigenstates constitute a complete
orthonormal basis of the space ${\cal H}$:
%
%
$\sum_{n=1}^{N} \ket{A_n} \bra{A_n} = 1, \bra{A_n} A_{n'} \rangle
= \delta_{nn'}  .$
%
%
The standard solution from linear algebra \cite{greub63} is to
write down the eigenvalue equation with respect to a given
orthonormal basis $\ket{k}, k=1,\ldots,N$, say. The $N^2$ matrix
elements ${\sf A}_{kk'} = \bra{k} {\widehat A} \ket{k'}$ determine
the operator ${\widehat A}$ uniquely and its eigenstates are
characterized by the coefficients $(\vec{A}_n)_k = A_{nk}$ in the
expansion $\ket{A_n} = \sum_{k} A_{nk} \ket{k}$. The number
$\lambda$ is an eigenvalue of $\widehat A$ if the characteristic
polynomial $P_A(\lambda)$ of the matrix  ${\sf A}$ vanishes,
%
%
$P_A(\lambda) = \det \left( {\sf A} - \lambda {\sf E} \right) = 0
,$
%
%
where ${\sf E}$ is the $(N\times N)$ unit matrix. Once the $N$
roots $A_n$ of the polynomial $P_A$ are known, the non-zero
solutions of the equation
\be
\left( {\sf A} - A_n \right) {\vec A}_n = 0 \, , \qquad n=1, \ldots , N\, ,
\label{eigenvec}
\ee
provide the eigenvectors $\ket{A_n}$ in the basis $\ket{k}$.
Analytic expressions for the eigenvalues $A_n$ in terms of the
elements of ${\sf A}$ exist only if $N\leq 4$. In general,
numerical methods are required to determine approximately the
roots of $P_A(\lambda)$.
%
%

The {\em quantum diagonalization} of hermitean matrices is based
on the assumption that the behaviour of a spin $s$ is described
correctly by non-relativistic quantum mechanics. This method will
make use of the `collapse of the wave function' as computational
resource. Note that the procedure does {\em not} depend on a
particular interpretation of quantum mechanics. Five steps are
necessary to achieve the diagonalization of a given matrix ${\sf
A}$ (supposed for simplicity not to have degenerate eigenvalues).
The individual steps will be described first in a condensed form;
subsequently, commentaries explain the technical details.

\begin{enumerate}
  \item \underline{Standard form of ${\sf A}$:}
      Write the hermitean $(N\times N$) matrix ${\sf A}$ as
      a combination of linearly
      independent hermitean \emph{multipole} operators
      ${\sf T}_\nu, \nu= 0, \dots ,N^2-1,$
\be
{\sf A} = \sum_{\nu=0}^{N^2-1} {\bf a}_{\nu} {\sf T}_{\nu} \, ,
\qquad {\bf a}_{\nu} = \frac{1}{N}\mbox{ Tr } \left[{\sf A} {\sf
T}_{\nu} \right] \in \mathbf{R} \, .
\label{expandgen} \ee
\item \underline{Identification of an observable:}
        Interpret the matrix ${\sf A}$ as an observable ${\sf H}_A$
        for a single quantum spin ${\sf S}$ with quantum number $s= (N-1)/2$,
\begin{equation}\label{observableha}
{\sf H}_A ({\sf S}) = \sum_{\nu=0}^{N^2-1} {\bf a}_{\nu} {\sf
T}_{\nu} ({\sf S}) \, ,
\end{equation}
using the expression of the multipoles ${\sf T}_{\nu} ({\sf S})$
in terms of the components of a spin.
\item \underline{Setting up a measuring device:}
  Construct an apparatus app(${\sf H}_A$) suitable to measure the
  observable ${\sf H}_A$.
\item\underline{Determination of the eigenvalues:}
Carry out measurements with the apparatus app(${\sf H}_A$) on a
spin $s$ prepared in a homogeneous mixture ${\hat \rho} = {\sf
I}/(2s+1)$. The output of each measurement will be one of the
eigenvalues $A_n$ of the matrix ${\sf A}.$ After sufficiently many
repetitions, all eigenvalues will be known.
\item \underline{Determination of the eigenstates:} Calculate the
eigenstates $\ket{A_n}$ of the matrix ${\sf A}$ on the basis of
Eq.\ (\ref{eigenvec}) and the experimentally determined eigenvalues
$A_n$. Alternatively, determine the eigenstates $\ket{A_n}$
\emph{experimentally} by methods of state reconstruction.
\end{enumerate}

Thus, the matrix ${\sf A}$ has been diagonalized without {\em
calculating} the zeroes of its characteristic polynomial by
traditional means. The fourth step solves the hard part of the
eigenvalue problem since it provides the eigenvalues $A_n$ of the
matrix ${\sf A}$. The comments to follow provide the background
necessary to perform the individual steps. Emphasis will be both
on the construction of a device measuring for a given hermitean
operator (Step 3) and on the working of a quantum mechanical
measurement (Step 4).

 \underline{Ad 1:}
The $N^2$ self-adjoint multipole operators ${\sf T}_{\nu} = {\sf
T}_{\nu}^\dag$ form a basis in the space of hermitean operators
acting on an $N$-dimensional Hilbert space ${\cal H}$
\cite{swift+80}. Two multipoles are orthogonal with respect to a
scalar product defined as the trace of their product: $(1/N)
\mbox{ Tr } \left[{\sf T}_{\nu} {\sf T}_{\nu'} \right] =
\delta_{\nu\nu'}.$

Consider now a Hilbert space ${\cal H}_s$ of dimension ($2s+1$)
which carries an irreducible representation of the group $SU(2)$
with the spin components $({\sf S_1},{\sf S_2},{\sf S_3})$ as
generators.  Then, the multipoles ${\sf T}_{\nu}, \nu = 1, \ldots,
N^2-1,$ are given by the symmetrized products ${\sf S}_{j_1} {\sf
S}_{j_2} \cdots {\sf S}_{j_a}, j_i = 1,2,3,$ and $a=0,1,\ldots,
2s,$ after subtracting off the trace (define ${\sf T}_0 \equiv
{\sf T}^{(0)} = {\sf E}$, the $(N\times N)$ unit matrix). The
index $a$ labels $(2s+1)$ classes with $(2a+1)$ elements
transforming among themselves under rotations; for the sake of
brevity, a collective index $\nu \equiv (a;j_1,\ldots,j_k)$ is
used. Explicitly, the lowest multipoles read
\be
 \quad {\sf T}^{(1)}_{j} = {\sf S}_j
\, , \quad {\sf T}^{(2)}_{j_1 j_2} = \frac{1}{2} \left( {\sf
S}_{j_1} {\sf S}_{j_2} +{\sf S}_{j_2}{\sf S}_{j_1} \right) -
\frac{\delta_{i_1 j_2}}{3} {\sf S}_{j_1}{\sf S}_{j_2} \, .
\label{mpoles012} \ee
The  set $\{ {\sf T}_\nu \} $ is a basis for the hermitean
operators on ${\cal H}_s$.

\underline{Ad 2:} Since the multipoles are expressed explicitly as
a function of the spin components not exceeding the power $2s$, it
is justified to consider them and, \emph{a fortiori}, the quantity
${\sf H}_A$ as an {\emph{observable} for a spin $s$.

\underline{Ad 3:} It is natural to expect that every self-adjoint
operator ${\widehat B}$ comes along with an apparatus
app(${\widehat B}$) capable of measuring it \cite{dirac58} . For
particle systems, setting up such a device remains a challenging task
for an experimenter.

For spin systems, the situation is different, however. Swift and
Wright \cite{swift+80} have shown  how to devise, in principle, a
{\em generalized Stern-Gerlach apparatus} which measures any
observable ${\sf H}_A({\sf S})$---just as a traditional
Stern-Gerlach apparatus measures the spin component ${\bf n} \cdot
{{\sf S}}$ along the direction ${\bf n}$. The construction
requires that arbitrary static electric and magnetic fields,
consistent with Maxwell's equations, can be created in the
laboratory. To construct an apparatus app$({\sf H}_A)$ means to
identify a spin Hamiltonian ${\sf H} ({\bf r}, {\sf S})$ which
splits an incoming beam of particles with spin $s$ into
subbeams corresponding to the eigenvalues $A_n$. The most general
Hamiltonian acting on the Hilbert space ${\cal H}$ of a spin $s$
reads
\be
{\sf H} ({\bf r}, {\sf S})
        = \sum_{\nu=0}^{N^2-1} \Phi_\nu ({\bf r})
                           {\sf T}_{\nu} \, ,
\label{genhamiltonian} \ee
with traceless (except for $\nu=0$) symmetric expansion
coefficients $\Phi_\nu ({\bf r}) (\equiv \Phi^{(k)}_{j_1 j_2
\ldots j_k}({\bf r}))$ which vary in space.  Tune the electric and
magnetic fields in such a way that the coefficients $\Phi_\nu({\bf
r})$ and its first derivative with respect to some spatial
direction, $r_1$, say,  satisfy
\be
\Phi_\nu ({\bf r} = 0) = \frac{\partial \Phi_\nu ({\bf r} =
0)}{\partial r_1} = {\sf a}_n \, .
\label{tune} \ee
This is always possible with realistic fields satisfying Maxwell's
equations. Then, the Hamiltonian in (\ref{genhamiltonian}) has two
important properties. (i) At the origin, ${\bf r} = 0$, it
coincides with the matrix  ${\sf H}_A$. (ii) Suppose that a beam
of particles with spin $s$ enters the generalized Stern-Gerlach
apparatus app(${\sf H}_A$) just described. At its center,
particles in an eigenstate $\ket{A_n}$, say, will experience a
force in the $r_1$ direction given (up to second order in distance
from the center) by
\be
F_1 ({\bf  r = 0}) = -\frac{\partial  \bra{A_n}{\sf H} ({\bf r} =
0, {\sf S}) \ket{A_n}} {\partial r_1} = - A_n \, , \qquad n= 1,
\ldots, 2s+1\, .
\label{separation} \ee
Consequently, particles with a spin projected onto one of the
eigenstates $\ket{A_n}$ of the operator ${\sf H}_A$ are separated
spatially by this apparatus. The procedure is entirely analogous
to that for a spin $1/2$ where a familiar Stern-Gerlach apparatus
is used (see \cite{swift+80} for details).

\underline{Ad 4:} The `projection postulate' of quantum mechanics
describes the effect of measuring an observable $\widehat B$ on a
system $\cal S$ by means of an apparatus app(${\widehat B}$). If the
system is prepared initially in a state with density matrix $\hat
\rho$ one has:
\be
\mbox{app}{({\widehat B})}: \quad \hat \rho \quad \stackrel{p_n}
{\longrightarrow} \quad \left( B_n ; {\hat \rho}_n \right) \, ,
\qquad p_n = \mbox{ Tr } \left[ \hat \rho {\hat \rho}_n \right] \,
.
\label{measure} \ee
The action of the apparatus is, with probability $p_n$, to throw
the system into an eigenstate ${\hat \rho}_n \equiv \ket{B_n}
\bra{B_n}$ of the observable $\widehat B$; the \emph{outcome} of
the measurement is given by the associated eigenvalue $B_n$. By
the way, the notion of `collapse' or `projection' can be avoided by
characterizing the process indirectly by refering to ``repeatable
measurements'' \cite{peres95}.

The outcome of an {\em individual} measurement cannot be predicted
due to the probabilistic character of quantum mechanics.
Therefore, the {\em probabilities} $p_n$, resulting from
(infinitely often) repeated measurements on identically prepared
systems, represent the essential link between theory and
experiment. They provide information about the state of the system
conditioned by the selected observable. Thus, a measurement
reveals (or confirms) properties of the {\em state} $\hat \rho$ of
the system while the observable ${\widehat B}$ at hand is assumed
to be known, including its eigenstates and eigenvalues. To put it
differently, the observable
 defines the {\em scope} of the possible results of a
measurement: the only possible outcomes are its eigenvalues $B_n$,
and, directly after the measurement the system necessarily resides
in the corresponding state $\ket{B_n}$.

As the occurrence of the eigenvalues is purely probabilistic, one
needs to repeat the experiment until all values $A_n$ have been
obtained. If the spin $s$ is prepared initially in a homogeneous
mixture, $\hat {\rho} = {\sf E}/(2s+1)$, the $(2s+1)$ possible
outcomes occur with equal probability. The probability not to have
obtained one specific value $A_n $ after $N_0 \gg N$ measurements
equals $1/{(2s+1)^{N_0}}$, decreasing exponentially with $N_0$.

\underline{Ad 5:} It would be very convenient now to `read out'
directly the quantum state ${\hat \rho}_n $ obtained from a single
measurement with result $A_n$. However, due to the no-cloning theorem
\cite{dieks82,wootters+82}, an unknown state cannot be
determined if only one copy of it is available. Upon repeating the
measurement a large number of times and keeping only those states
with the {\em same} eigenvalue $A_n$, one produces an {\em
ensemble} of systems prepared identically in the state ${\hat
\rho}_n$. This is sufficient to reconstruct an unknown state since
a density matrix $\hat \rho$ can be written as
\begin{equation}
{\hat {\rho}} =
    \frac{1}{N} \sum_{\mu=1}^{N^2}  P_{\mu}
    {\widehat Q}^{\mu} \, , \qquad N = 2s+1\, ,
\label{expanddensity}
\end{equation}
where the coefficient $ P_{\mu} \equiv  \bra{{\bf n}_{\mu}} \hat
\rho \ket{{\bf n}_{\mu}}$ is the probability to find the system in
a coherent spin state $\ket{{\bf n}_{\mu}}$. The operators
${\widehat {Q}}^{\mu}, \mu = 1, \ldots, N^2$, form a basis for
hermitian operators, similar to but different from the multipoles
${\sf T}{_\nu}$ \cite{weigert99/1}. Thus, Eq.\
(\ref{expanddensity}) parametrizes $\hat \rho$ by expectation
values $ P_{\mu}$  which can be measured by a standard
Stern-Gerlach apparatus.-
%
%
%

%
%
In sum, the basic ingredient of quantum diagonalization is the
`collapse' of the wave function projecting any state onto a
randomly selected eigenstate of the measured observable.
Generalizations of this approach are expected to include the
diagonalization of unitary matrices and the determination of roots
of polynomials.

Usually, a measurement is thought to confirm or reveal some
information about the state of the system. Here, on the contrary,
the idea is to learn something about the measured observable
instead. Why is this possible? It is fundamental to realize that
the {\em input} required to actually measure $\widehat A$ differs
from the {\em output} of the experiment: for a measurement of
$\widehat A$, the construction of an apparatus app(${\widehat A}$)
is sufficient which is \emph{possible without}knowing eigenvalues
and eigenstates of $\widehat A$}. Necessarily, after a measurement
partial information about the spectral properties of the
observable $\widehat A$ is available according to (\ref{measure}).
This is due to the constraints (i) that the possible outcomes of
measuring ${\widehat A}$ are its eigenvalues and (ii) that the
system subsequently will occupy the corresponding eigenstate.
Thus, if the eigenstates and eigenvalues of ${\widehat A}$ not
known initially, one indeed acquires information about them by measuring
${\widehat A}$.

The quantum mechanical diagonalization appears to be neither an
analog nor a digital calculation. It is not based on the
representation of a mathematical equation in terms of a physical
system which then would `simulate' it. Similarly, no `software
program' is executed which would implement an diagonalization
algorithm. One might best describe the measuring device
app(${\sf H}_A$)  as a `special purpose machine' based on the
projection postulate.

For the time being, the method introduced here is important from a
conceptual but not a technological point of view. On the one hand,
the diagonalization of matrices is not a hard problem  such as
factorization of large integer numbers; on the other, the actual
implementation in the laboratory is challenging. It is important,
however, that there is no physical principle which would forbid
the construction of such a machine. Further, it is expected to be
fruitful from a conceptual point of view since it provides a
different perspective on the projection postulate \cite{jammer74}.
Quantum diagonalization as introduced here shows that---in an
unexpected way---standard quantum mechanics attributes {\em
computational power} to the measurement of an observable. The fact
that one can use a measurement to perform calculations might turn
into an argument in favor of the `reality' of the quantum
mechanical projection postulate.


\begin{thebibliography}{10}

\bibitem{bennet+84}
C.~H. Bennet and G.~Brassard: \emph{Quantum cryptography: Public
key distribution and coin tossing.} In: Proceedings of IEEE
International Conference on Computers, Systems, and Signal
Processing,  175 (1984).

\bibitem{schroedinger35/2}
E.~Schr\"odinger: \emph{Die gegenw\"artige Situation in der
Quantenmechanik.} Naturwissenschaften,  {\bf 23}, 823 (1935).

\bibitem{shor94}
P.~W. Shor: \emph{Algorithms for quantum computation: discrete log
and factoring}. In: {\em Proceedings of the 35th Symposium on
  the Foundations of Computer Science}, edited by S.~Goldwasser.
   IEEE Computer Society Press, 124 (1994).

\bibitem{shor95}
P.~W. Shor: \emph{Scheme for reducing decoherence in quantum
computer memory}, Phys.\ Rev.\ A, {\bf 52} R2493 (1995).

\bibitem{steane96}
A.~M. Steane: \emph{Error correcting codes in quantum theory},
Phys.\ Rev.\ Lett.\ {\bf 77} 793 (1996).

\bibitem{greub63}
W.~H. Greub: {\em Linear Algebra} (Springer, Berlin, 1963).

\bibitem{swift+80}
A.R. Swift and R.~Wright: \emph{Generalized Stern-Gerlach
experiments and the observability of arbitrary operators}, J.\
Math.\ Phys.\ {\bf 21}, 77 (1977).

\bibitem{dirac58}
P.~A.~M. Dirac: {\em The Principles of Quantum Mechanics}, (Oxford
University Press, Oxford, 1958).

\bibitem{peres95}
A.~Peres: {\em Quantum theory: concepts and methods}, (Kluwer
Academic Publications, 1995).

\bibitem{dieks82}
D.~Dieks: {\em Communication by epr-devices}, Phys.\ Lett.\ A {\bf
92}, 271 (1982).

\bibitem{wootters+82}
W.~K. Wootters and W.~H. Zurek: \emph{A single quantum cannot be
cloned}, Nature {\bf 299}, 80 (1982).

\bibitem{weigert99/1}
St. Weigert: {\em Quantum time evolution in terms of nonredundant
probabilities}, Phys.\ Rev.\ Lett.\ (in print, 2000)
(=quant-ph/99030103).

\bibitem{jammer74}
M.~Jammer: {\em Conceptual Foundations of Quantum Mechanics}.
\newblock (Wiley, New-York, 1974).

\end{thebibliography}
\end{document}